 \documentclass[1.5p,times]{elsart}
\usepackage{natbib}
\usepackage{graphicx}
\usepackage{color}
 \usepackage{epsfig}
\usepackage{amssymb}
\usepackage{amsmath}
\usepackage{listings}
\begin{document}

\begin{frontmatter}



\title{Absolute properties of the neglected eclipsing B-type binary HD 194495\thanksref{italy}}


\author[OCakirli]{\"{O}m\"{u}r \c{C}ak{\i}rl{\i}}, 
\author[OCakirli]{Esin Sİpahi} \&
\author[OCakirli]{Cafer {\.I}bano\v{g}lu}

\address[OCakirli]{Ege University, Science Faculty, Astronomy and Space Sciences Dept., 35100 Bornova, \.{I}zmir, Turkey. $e-mail$: omur.cakirli@ege.edu.tr}
\address[italy]{Based on observations collected at Catania Astrophysical Observatory (Italy)}

\begin{abstract}
We present the results of the high-resolution spectroscopic observations of the neglected binary system HD\,194495 (B3\,IV-V+B4\,V). A combined
analysis of three different photometric data set ($Tycho$ B$_T$ and V$_T$ photometry, H$_p$-band data of $Hipparcos$ and $V-$band
data of ASAS3 photometry) and radial velocities indicates that the system has an orbital period of 4.90494 $\pm$ 0.00005 days and an
inclination of 69$\pm$1 degrees. This solution yields masses and radii of $M_{\rm 1}$ = 7.57$\pm$0.08 $M_{\odot}$ and
$R_{\rm 1}$ = 5.82$\pm$0.03 $R_{\odot}$ for the primary and $M_{\rm 2}$ = 5.46$\pm$0.09 $M_{\odot}$ and $R_{\rm 2}$ = 3.14$\pm$0.08
$R_{\odot}$ for the secondary. Based on the position of the two stars plotted on a theoretical H-R diagram, we find that the age
of the system is $\gtrsim$ 28 Myr, according to stellar evolutionary models. The spectroscopic and photometric results are in agreement
with those obtained using theoretical predictions.
\end{abstract}

\begin{keyword}
Stars: binaries; Eclipsing — stars: fundamental parameters; Individual — Method: Spectroscopy; stars: HD\,194495
\end{keyword}

\end{frontmatter}

\section{Introduction}
The masses and radii of stars are two fundamental parameters in stellar astrophysics. Accurate mass determinations of B--type stars 
still are urgently required, since actual masses are based on a very small number of eclipsing binaries. Data of this kind are important for 
tests of stellar models, with far-reaching implications such as modelling of stellar components of galaxies. 
Our main motivations to study this system to explain the nature of the B-star binary HD\,194495. This effort is part of an ongoing research 
to determine the masses and radii of such systems.

The star HD\,194495 was classified as B7\,V single-lined, massive spectroscopic binary by \citet{Monet79}. System consists of an 
evolved, more massive, and more luminous primary component and a main sequence secondary star. It has an eccentric orbit ($e$=0.14) 
and an orbital period of 4.90 days. Several radial velocity measurements of primary component of HD\,194495 by \citet{Monet79} show 
range from -80 km\,s$^{-1}$ to 100 km\,s$^{-1}$, but no further investigations were made into this probable velocity variable and 
radial velocity of the second component measurements. The eclipsing nature of its light curve was first noticed with the $Hipparcos$ 
satellite (HIP\,100719) and later $ASAS$ photometric survey (ASAS\,202511+2129.3) where it was classified as an $unsolved~variables$ 
eclipsing binary and known as V399\,Vulpecula \citep{Kazo99}.

As one of only a few eclipsing B-star binary HD\,194495 provides a potentially important system because it is a test case for stellar 
structure and evolution models. It is critical that well-determined values for the current separation and component masses be 
determined. In this paper, we use the optical spectra of HD\,194495 to reveal the nature of its light variability and physical 
properties in combination with the photometric data obtained by Hipparcos and ASAS-3 as well as with the our reconstruction of 
spectra. The paper is organized as follows. In \S 2 the spectroscopic observations, data analysis, temperature estimations, 
reddening and reconstruction of spectra are described. Derived absolute parameters of the stars from the combination spectroscopic 
and photometric results are given in \S 3. We compare the individual masses based on the spectroscopic and photometric orbits 
to those predicted by theoretical methods and discuss their implications in \S 4.

\section{Spectroscopic observations}
Spectroscopic observations have been performed with the \'{e}chelle spectrograph (FRESCO) at the 91-cm telescope of Catania Astrophysical 
Observatory. The spectrograph is fed by the telescope through an optical fibre ($UV$--$NIR$, 100 $\mu$m core diameter) and is located, in 
a stable position, in the room below the dome level. Spectra were recorded on a CCD camera equipped with a thinned back--illuminated SITe 
CCD of 1k$\times$1k pixels (size 24$\times$24 $\mu$m). The cross-dispersed \'{e}chelle configuration yields a resolving power 
$R$=$\lambda/\delta\lambda$=22\,000, as deduced from the full width at half maximum of the lines of the Th--Ar calibration lamp. The 
spectra cover the wavelength range from 4300 to 6650 {\AA}, split into 19 orders. In this spectral region, and in particular in the 
blue portion of the spectrum, there are several lines useful for the measure of radial velocity, as well as for spectral classification 
of the stars.

The data reduction was performed by using the \'{e}chelle task of IRAF\footnote{IRAF is distributed by the National Optical Observatory, which 
is operated by the Association of the Universities for Research in Astronomy, inc. (AURA) under cooperative agreement with the National Science 
Foundation} package following the standard steps: background subtraction, division by a flat field spectrum given by a halogen lamp, wavelength 
calibration using the emission lines of a Th-Ar lamp, and normalization to the continuum through a polynomial fit. 

Seventeen spectra of HD\,194495 were collected during the 20 observing nights between August 14 and September 22, 2007. Typical exposure times for 
the HD\,194495 spectroscopic observations were between 2400 and 2600 s. The signal-to-noise ratio ($S/N$) achieved was between 70 and 120, depending 
on atmospheric condition. $\alpha$ Lyr (A0V), 59 Her (A3IV), $\iota$ Psc (F7V), HD 27962 (A2IV), and $\tau$ Her (B5IV) were observed during each run 
as radial velocity and/or rotational velocity templates. The average $S/N$ at continuum in the spectral region of interest was 150--200 for the standard 
stars.

\begin{figure*}
  \begin{center}
    \includegraphics[scale=0.75]{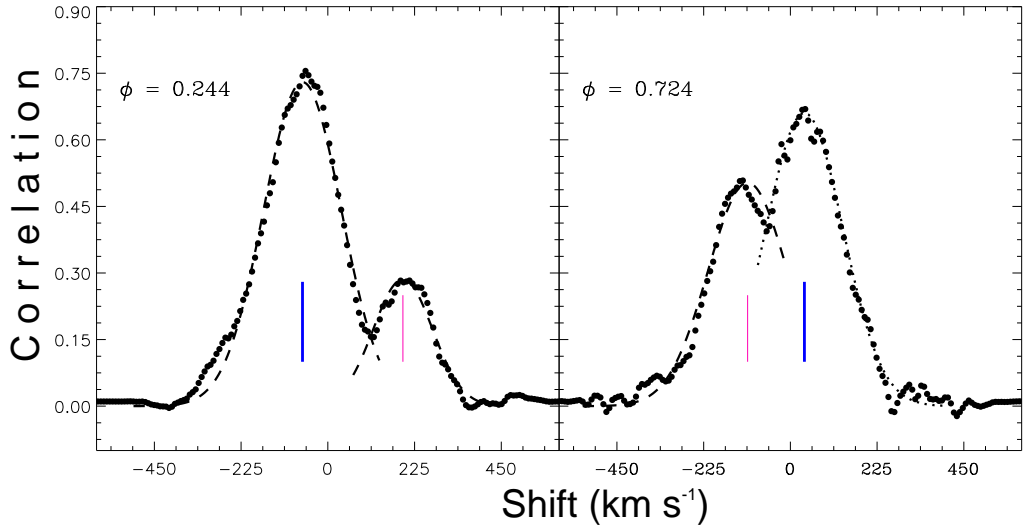}
  \end{center}
  \caption{Sample of Cross Correlation Functions between HD\,194495 and the radial velocity template spectrum in
first and second quadrature phase. The horizontal axis is relative radial velocities, and vertical axis is normalized 
cross-correlation amplitude. Note that splittings at stronger peaks. The phases of observations and 
the wavelength of the peak centers of the primary (thick bar) and secondary (thin bar) component are also marked.
}  \label{fig:ccf}
\end{figure*}

\begin{table}
\centering
\caption{Radial velocities of the components of HD\,194495. The columns give the heliocentric Julian date, the
orbital phase, the radial velocities of the two components with the corresponding errors, and the average S/N of the spectrum. }
\begin{tabular}{@{}ccrcccc@{}}
\hline
\textsf {HJD} & Phase& \multicolumn{2}{c}{Star 1 }& \multicolumn{2}{c}{Star 2 } &  $<S/N>$ \\
  2\,454\,300+ &  & \textsf{{\bf V$_p$}} & $\sigma$ & \textsf{{\bf V$_s$}} & $\sigma$&  \\
\hline
27.4695   &   0.0500 &  -54 &	 4.1&	 --  &      -- &     89        \\
28.4235   &   0.2445 & -144 &	 1.2&	 160 &	   4.6 &    120$^a$    \\
29.3672   &   0.4369 &  -35 &	 5.0&	 --  &	    -- &     97        \\
30.4330   &   0.6542 &	 80 &	 1.6&	-140 &	   9.1 &    106        \\
31.3786   &   0.8470 &	 74 &	 1.5&	-137 &	   9.9 &     95        \\
33.4202   &   0.2632 & -138 &	 3.1&	 151 &	   5.1 &    106$^a$    \\
35.3825   &   0.6633 &	 78 &	 1.2&	-134 &	   9.2 &     80        \\
36.4028   &   0.8713 &	 65 &	 3.2&	-127 &	   8.1 &     90$^a$    \\
37.3967   &   0.0739 &  -73 &	 4.2&	 --  &	    -- &     90        \\
38.3786   &   0.2741 & -136 &	 3.2&	 151 &	   4.4 &    113$^a$    \\
43.3666   &   0.2910 & -127 &	 3.5&	 132 &	   7.6 &     95        \\
46.4666   &   0.9231 &	 40 &	 2.9&	 -87 &	  11.1 &     97        \\
48.3967   &   0.3165 & -110 &	 2.1&	 124 &	   7.7 &    111$^a$    \\
50.3967   &   0.7243 &	 90 &	 3.0&	-157 &	  10.1 &    120$^a$    \\
64.4143   &   0.5822 &	 56 &	 1.5&	-105 &	  15.8 &     72        \\
65.4489   &   0.7931 &	 89 &	 2.1&	-152 &	   4.1 &    115$^a$    \\
66.4121   &   0.9895 &	 -6 &	 6.2&	  -- &	    -- &     90        \\
\hline
\end{tabular}
\begin{list}{}{}
\item[$^a$]{\small Used also for rotational velocities ($v\sin i$) measurements.}
\end{list}
\end{table}

\subsection{Spectroscopic analysis}
The radial velocities of HD\,194495 were obtained by cross--correlating of \'{e}chelle orders of 
HD\,194495 spectra with the spectra of the bright radial velocity standard stars $\alpha$ Lyr and $\tau$ Her
\citep{Nord63}. For this purpose the IRAF task \textsf{fxcor} was used. 

Figure 1 shows example of CCF at first and second quadrature phase. The two peaks correspond 
to each component of HD\,194495. The stronger peaks in each CCF correspond to the more luminous component that 
have a larger weight into the observed spectrum. We adopted a two-Gaussian fit algorithm to resolve cross-correlation 
peaks near the second quadrature when spectral lines are visible separately. At this phase, absorption 
lines of the primary and secondary components of the system can be easily recognized in the range between 
4300-6800 \AA. These regions include the following lines: He\,{\sc i} 4387 \AA, Mg\,{\sc ii} 4481 \AA, He\,{\sc i} 
4713, He\,{\sc i} 5016 \AA, He\,{\sc i} 4917 \AA, He\,{\sc i} 5876 \AA. We limited our analysis to the echelle orders 
in the spectral domains in the range, which include several photospheric absorption lines. We have disregarded 
very broad lines like H$_{\alpha}$, H$_{\beta}$ and H$_{\gamma}$ because their broad wings affect the CCF and lead 
to large errors. A double-lined Gaussian fit was used to disentangle the CCF peaks and determine the RVs of each 
component. Following the method proposed by \citet{Pen01} we first made two-Gaussian fits of the well 
separated CCFs using the deblending procedure in the IRAF routine {\sf splot}. The average fitted FWHM is 234$\pm$8 and
178$\pm$9 km\,s$^{-1}$ for the primary and secondary components, respectively. In Figure 1 we show a sample of double-Gaussian 
fit. Indeed, the shapes and velocities corresponding to the peaks of the CCFs are slightly changed. By measuring 
the areas enclosed by the Lorentzian profiles of the spectral lines belonging to the primary (A$_1$) and secondary (A$_2$) 
we estimate the light ratio of the primary star (F$_1$) to the secondary (F$_2$). In this way, we were able to obtain an estimate of 
the monochromatic flux ratio in the red and blue part of the spectra of F$_1$/F$_2 \sim $A$_1$/A$_2$=0.63 based upon the relative 
line depths of the spectral components.

\subsubsection{Resulting radial velocities}
The resulting radial velocities are listed in Table 1 together with their standard errors. The observational
points and their error bars are displayed in Figure 2 as a function of orbital phase as calculated by means of
the ephemeris (equation 1) and fixed it during the orbital solutions. Other parameters, such as the velocity semi-amplitude
of the components (K$_{1,2}$), systemic velocity (V$_\gamma$), longitude of periastron ($\omega$), orbital eccentricity
($e$), and time of periastron passage (T$_0$) were converged. The final solution gave K$_1$=116$\pm$4 km s$^{-1}$,  
K$_2$=161$\pm$6 km s$^{-1}$, V$\gamma$=-15$\pm$1 km s$^{-1}$, $\omega$=2.72$\pm$0.09, and $e$=0.12$\pm$0.07. Our 
spectroscopic mass ratio is $q$=M$_2$/M$_1$=0.72. Hence, our analysis gives the following parameters: 
$M_1$ $\sin^3 i$=6.16$\pm$0.05 M$_{\odot}$, $M_2$ $\sin^3 i$=4.44$\pm$0.06 M$_{\odot}$, and $a$ $\sin i$=26.68$\pm$0.01 R$_{\odot}$.

\begin{figure}
  \begin{center}
    \includegraphics[scale=0.48]{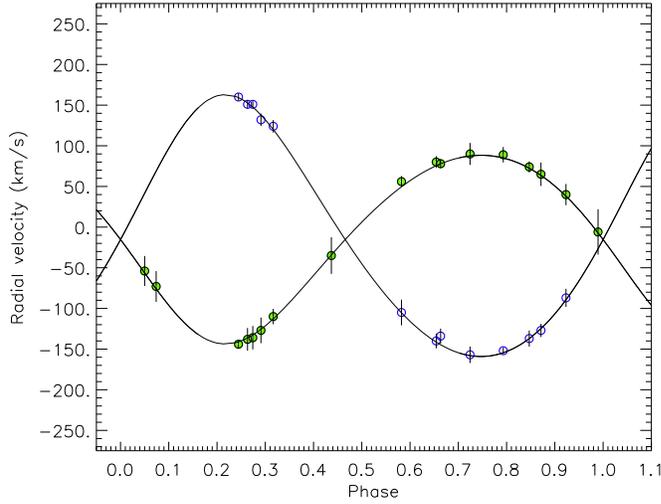}
  \end{center}
  \caption{Radial velocity curve folded on a period of 4.904938 days. Points with error bars (error bars are masked by 
the symbol size in some cases) show the radial velocity measurements for the components of the system 
(primary: filled circles, secondary: open circles). 
}  \label{fig:RV}
\end{figure}

\subsection{Spectral classification and temperature estimates}
We have used the spectra to reveal the spectral type of the primary component of HD\,194495. For this 
purpose we have measured the equivalent widths ($EW$) of photospheric absorption lines for the spectral classification (Table\,2). 
We have followed the procedures described by \citet{Hernandez04}, choosing hydrogen and helium lines in the whole wavelength 
region, where the contribution of the primary component is considerably larger than that of the secondary in the spectra.
From the calibration relations of $EW$--Spectral-type given by \citet{Hernandez04}, we have derived a 
spectral type of B3$\pm$1 for the primary  component. The effective temperature deduced from the calibrations 
of \citet{Dri00}, and \citet{JagerNie87} is 18\,600$\pm$650 and 19\,100$\pm$680\,K 
for the primary component, respectively. The mean effective temperature of the primary component deduced from 
the spectra is therefore 19\,000$\pm$550\,K . 

\begin{table}
\begin{center}
\caption{Equivalent widths of the selected lines in the spectra.}
\label{Table 1.}
\begin{tabular}{lc}
\hline
 $Spectral~lines$ & EW$_{primary}$ (\AA)  \\
\hline
He \sc i + F$e$ \sc i $\lambda$ 4922				&0.26$\pm$0.11 		\\
H$\gamma$  	$\lambda$ 	   4349						&0.37$\pm$0.11 		\\
He \sc i $\lambda$ 		   5876						&0.35$\pm$0.13 		\\
He {\sc i}+Fe {\sc i} $\lambda$ 4387				&0.35$\pm$0.21 		\\
H$\beta$ 				4861						&5.31$\pm$0.42 		\\
\hline \\
\end{tabular}
\end{center}
\end{table}

HD\,194495 is listed in several large photometric databases consolidated in the $Hipparcos$ Catalogue, which 
provide optical magnitudes of B=7$^m$.09$\pm$0$^m$.02, V=7$^m$.07$\pm$0$^m$.01. Since the magnitudes collected 
from photometric measurements and the colors are inconsistent no attempt has been made to calculate the 
effective temperatures of the components. The infrared magnitudes are taken from 2MASS \citep{Cutri03} 
catalog as J=7$^m$.159$\pm$0$^m$.009, H=7$^m$.221$\pm$0$^m$.017, and K=7$^m$.249$\pm$0$^m$.011.
The observed infrared colours of J-H=-0$^m$.162$\pm$0$^m$.031 and H-K=-0$^m$.028$\pm$0$^m$.018  
correspond to a combined spectral type of B2$\pm$2 is an agreement with that we derived by spectral 
lines alone. Hence V-K, J-H and H-K colors of the primary component corresponds to a spectral type of 
B2$\pm$1.

\subsection{Reddening}
The measurement of reddening is a key step in determining the absolute temperature scale (and therefore the distance) of 
eclipsing binaries. In addition to moderate distance determined by the Hipparcos mission, some reddening is expected for 
HD\,194495 due to its low galactic latitude ($l$=63$^{\circ}.03$, $b=-9^{\circ}.34$).

Our spectra cover the interstellar Na{\sc i} (5890 and 5896 \AA) doublets, which is excellent estimators of the reddening as 
demonstrated by \citet{MunariZwit97}. They calibrated a tight relation linking the Na {\sc i} D2 (5890 \AA) and K{\sc i} 
(7699 \AA) equivalent widths with the E(B-V) reddening. On spectra obtained at quadratures, lines from both components 
are un-blended with the interstellar ones, which can therefore be accurately measured. We derive an equivalent width of 
0.21$\pm$0.03 \AA~for only Na{\sc i}, which corresponds to E(B-V)= 0$^m$.08$\pm$0$^m$.02. K{\sc i} interstellar line is 
out of our spectral range as given in wavelength region in previous section.

\subsection{Rotational velocity and reconstruction of spectra}
The width of the cross-correlation profile is a good tool for the measurement of $v \sin i$ (see, e.g., 
\citet{Que98}). The rotational velocities ($v \sin i$) of the two components were obtained by 
measuring the FWHM of the CCF peaks in ten high-S/N spectra of HD\,194495 acquired close to the 
quadratures, where the spectral lines have the largest Doppler-shifts. In order to construct a 
calibration curve FWHM--$v \sin i$, we have used an average spectrum of HD~27962, acquired with 
the same instrumentation. Since the rotational velocity of HD~27962 is very low but not zero 
($v \sin i$ $\simeq$11 km s$^{-1}$, e.g., \citet{Royer02}), it could be 
considered as a useful template rotating faster than $v \sin i$ $\simeq$ 10 km s$^{-1}$. The spectrum of 
HD~27962 was synthetically broadened by convolution with rotational 
profiles of increasing $v \sin i$ in steps of 5 km s$^{-1}$ and the cross-correlation with the original 
one was performed at each step. The FWHM of the CCF peak was measured and the FWHM-$v \sin i$ 
calibration was established. The $v \sin i$ values of the components of HD\,194495 were derived 
from the FWHM of their CCF peak and the aforementioned calibration relations, for a few wavelength 
regions and for the best spectra. This gave values of 61$\pm$5 km s$^{-1}$ for the primary star 
and 39$\pm$7 km s$^{-1}$ for the secondary star. 

\begin{figure}
  \begin{center}
    \includegraphics[scale=0.65]{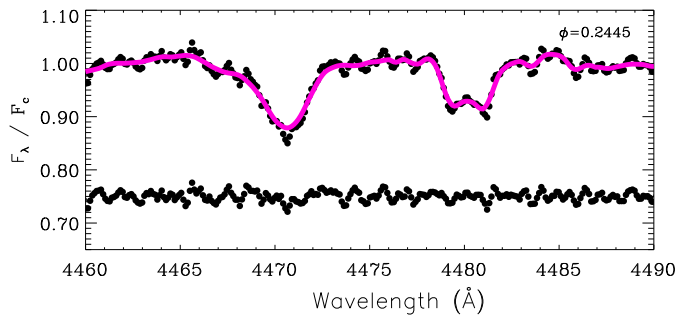}
  \end{center}
  \caption{Observed spectrum of HD\,194495 (large dots) in the Mg\,{\sc ii} $\lambda$4481. The synthetic spectrum (B3V+B4V) is 
displayed with continuous line in the same boxes. The differences (observed-synthetic, shifted) are plotted in the bottom 
of panel.
}  \label{fig:V_rot}
\end{figure}

We performed an accurate spectral classification and measured rotational velocities of the components in order to
search for the best combination of the two standard-star spectra able to reproduce the observed spectrum of the 
HD\,194495. From the observed two standard star spectra of, acquired with the same instrumentation, $\tau$\,Her and $\alpha$\,Lyr 
are used as standard stars for the primary and secondary, respectively. For the construction of the reproduced observed 
spectrum of the system the spectra of the $\tau$\,Her (B5IV) and $\alpha$\,Lyr (A0V) have been rotationally broadened by convolution with 
the appropriate rotational profile and then have co-added, properly weighted by using physical parameters
($T_1$, $T_2$, $R_1$, $R_2$, $v_{1,2}sini$) of the components as input parameters and Doppler-shifted according to the radial velocity 
solution derived in next section. 

Figure 3 shows the reconstructed spectra for the primary and secondary with the best fit spectra overplotted. The relative depths of 
He\,{\sc i}  $\lambda$4471 and Mg\,{\sc ii} $\lambda$4481 are good temperature indicators throughout the B-star sequence \citep{Will09} 
plus the H$_\gamma$ line that is sensitive to gravity (linear stark effect). Specifically, the He\,{\sc i}  $\lambda$4471 line 
gets weaker while the Mg\,{\sc ii} $\lambda$4481 line gets stronger as temperature decreases.

The resolving power of the instrument is unsuitable for attempting classical spectral typing. However, we can arrive 
at an estimate of the spectral types for each components in the HD\,194495 system by comparing the derived effective temperatures 
in \S 2.2 and \S 3. According to the Table 2 of \citet{Bohm81} the effective temperature and gravity of the
primary of HD\,194495 is most consistent with a B3\,IV-V star while the secondary matches most closely with a B4\,V star, and these 
classifications are listed in Table 3.

\begin{table}
\caption{Spectroscopic reconstruction parameters of the components.}
\label{Table 3.}
\begin{tabular}{lcc}
\hline
 Parameter & Primary & Secondary \\
\hline
Spectral Type					&B3\,IV-V$^a$								&B4\,V$^a$			\\
T$_{eff}$ (K)					&19\,000$\pm$320 							&18\,250$\pm$520	\\
log $g$ (cgs)					&3.79$\pm$0.25 								&4.18$\pm$0.25		\\
$vsini$	(km s$^{-1}$)			&61$\pm$5 									&39$\pm$7			\\
$F_1/F_2$ 						& \multicolumn{2}{c}{0.63$\pm$0.12}								\\
\hline \\
\end{tabular}
\begin{list}{}{}
\item[$^a$]{\small The spectral types are estimated from derived values of T$_{eff}$ and log $g$.}
\end{list}
\end{table}

\section{Combined radial velocity and light curve solution}
\subsection{The binary ephemeris}
Studying of the light curves was achieved on the basis of three different photometric data sets [Tycho B$_T$ and V$_T$ photometry, 
H$_p$-band data of Hipparcos and V-band data of ASAS3 \citep{Poj02}]. Meaningfully photometric observations of 
HD\,194495 were made by the Hipparcos mission and 78 H$_p$ magnitudes were listed by \citet{vanlee07}. These magnitudes 
were obtained in a time interval of about three years. The accuracy of the Hipparcos data is about $\sigma_{H_p}$ $\sim$ 0.01. These 
measurements  are plotted against the orbital phase in the bottom panel of Figure 4. In spite of their low precision, the Hipparcos and
Tycho data clearly show light variation when the observations are phased with the ephemeris, limited by the lack of observations in the
primary and secondary minima and by insufficient sensitivity.

\begin{figure}
  \begin{center}
    \includegraphics[scale=0.50]{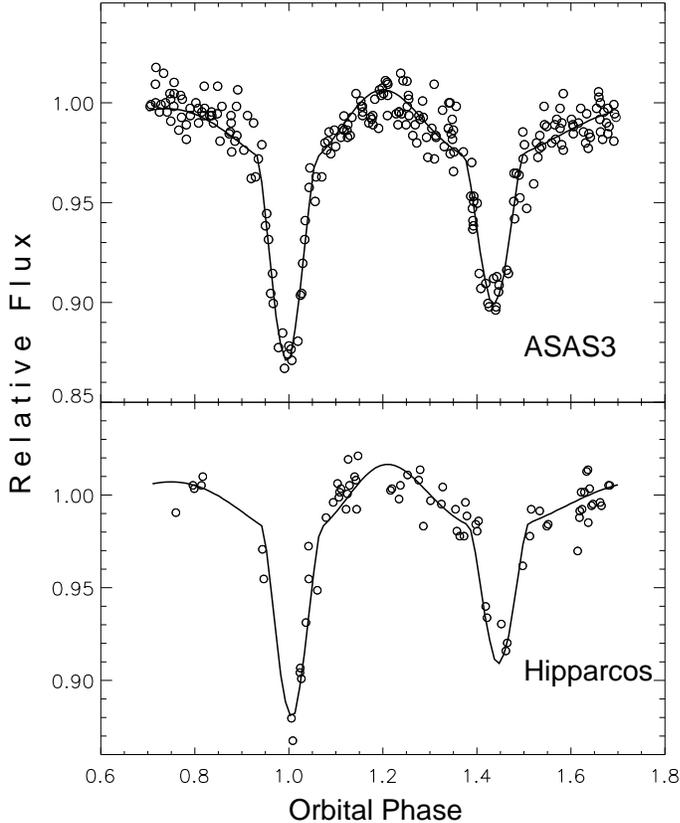}
  \end{center}
  \caption{The ASAS3-V band and H$_p$ band light curves of HD\,194495 with the best fit model overlaid as shown in Table\,5.
}  \label{fig:LC}
\end{figure}

The All Sky Automated Survey database contains light curves for $\sim$39\,000 previously 
unknown variable stars. We extracted the V-band light curve for HD\,194495 ($ASAS$ 202511+2129.3) from this catalog, removing 
points deemed lower quality by the data reduction pipeline used by the ASAS. The 238 photometric measurements of 
ASAS, including eclipses, were used to determine the light-curve elements. All available photometric data are phased 
and re-scaled in Figure 4. It is clear from Figure 4 that HD\,194495 is an eccentric binary, with the secondary eclipse at 
phase 0.44 relative to the primary eclipse at phase 0.00. The scatter of the data in the out-of-eclipse phases 
was about $0^{\rm m}.03$.

The ASAS photometry permits determination of only one seasonal moment of primary eclipse, but this enable us to obtain an improved
primary eclipse ephemeris. We determine from the data an ephemeris of:
\begin{equation}
Min I(HJD)=2\,452\,860.6478(25)+4^d.904938(33) \times E
\end{equation}
where the errors in the last significant digits are given in parentheses.

\subsection{Determination of the photometric elements}
In order to reproduce the observed characteristics of the photometric light curves, we 
analysed them with the Wilson--Devinney code implemented into the PHOEBE package tool by \citet{PrsaZt05}
for the LC and differential correction (DC) fits. For the analysis, combined light curves of three data set
was modelled. Considering the spectroscopic analysis the temperature of the primary was fixed to 
19\,000 K, and the bolometric albedo and gravity brightening coefficients were set to unity, as 
generally found for stars with the radiative envelopes \citep{Zeipel24}. The logarithmic limb 
darkening law was used and limb-darkening coefficients were taken from \citet{vanham03}. The 
surface potential ($\Omega_{1,2}$), light factors of the components ($l_{1,2}$) and orbital 
inclination ($i$) were adjustable parameters  during the light-curve modelling.

\begin{table}
  \caption{Results from the simultaneous solution of ASAS$_V$ and Hipparcos$_{V_p}$ band light curves of HD\,194495.}
  \label{parameters}
  \begin{tabular}{lcrr}
  \hline
   Parameter & ASAS+H$_p$&&\\
   \hline
   $i (^{\circ})$ 					& 69$\pm$1		  	      \\
   $T_{1}$ (K) 						& 19\,000[\textsf{Fix}]   \\
   $T_{2}$ (K)						& 17\,900$\pm$70 	      \\
   $\Omega_{1}$ 					& 5.690$\pm$0.038 	      \\
   $\Omega_{2}$ 					& 7.054$\pm$0.057 	      \\
   $q_{spec}$ 						& 0.723[\textsf{Fix}]	  \\
   $\dfrac{ L_{1_{Hipp.}}} {(L_{1+2})_V}$ 	& 0.609$\pm$0.091  	  	  \\
   $\dfrac{L_{1_{ASAS3}}} {(L_{1+2})_V}$ 		& 0.630$\pm$0.091  	  	  \\
   $r_1$							& 0.2035$\pm$0.0013  	  \\
   $r_2$							& 0.1123$\pm$0.0017  	  \\
   $\chi^2$							& 0.0211  		      	  \\
  \hline
  \end{tabular}
\end{table}

The mass ratio, $q$=M$_2$/M$_1$, is very important parameter in the light curve analysis, because the 
WD code is based on Roche geometry which is sensitive to this quantity. The mass ratio, eccentricity and 
longitude of periastron were determined from the radial velocity analysis was kept as a fix value. The 
iterations were carried out automatically until convergence and a solution was defined as the set of 
parameters for which the differential corrections were smaller than the probable errors. The combined 
light curve was analized and the weighted means of the parameters $i$, T$_2$, $\Omega_1$, 
$\Omega_2$, $r_1$ and $r_2$ were computed. Our final results are listed in Table 4 and the computed light 
curves are shown as continuous lines in Figure 4. The uncertainities assigned to the adjusted parameters 
are the internal errors provided directly by the Wilson-Devinney code.

\begin{table}
  \caption{Fundamental parameters of HD\,194495.}
  \label{parameters}
  \begin{tabular}{lcc}
  \hline
  & \multicolumn{2}{c}{HD\,194495} 		\\
   Parameter 																					 & Primary											&	Secondary													\\
   \hline
   $a$ (R$_{\odot}$)																		 &\multicolumn{2}{c}{28.58$\pm$1.12}																	\\
   $V_{\gamma}$ (km s$^{-1}$)													 &\multicolumn{2}{c}{-15$\pm$1} 																			\\
   $q$																								 &\multicolumn{2}{c}{0.723$\pm$0.003}																\\
   Mass (M$_{\odot}$) 																	 & 7.57$\pm$0.08 								&5.46$\pm$0.03												\\
   Radius (R$_{\odot}$) 																	 & 5.82$\pm$0.03 								&3.14$\pm$0.08												\\
   $\log~g$ ($cgs$) 																			 & 3.50$\pm$0.02 								&3.54$\pm$0.03												\\
   $T_{eff}$ (K)																				 & 19\,000$\pm$550							&17\,800$\pm$600      									\\
   $(vsin~i)_{obs}$ (km s$^{-1}$)													 & 61$\pm$2										&39$\pm$4       												\\
   $(vsin~i)_{calc.}$ (km s$^{-1}$)												 & 60$\pm$1										&32$\pm$1		    											\\
   $\log~(L/L_{\odot})$																	 & 3.93$\pm$0.06								&3.64$\pm$0.10       										\\
   $d$ (pc)																						 & \multicolumn{2}{c}{274$\pm$5}																			\\
   $J$, $H$, $K_s$ (mag)$^{*}$														 & \multicolumn{2}{c}{7.159$\pm$0.018, 7.221$\pm$0.027, 7.249$\pm$0.027}	\\
$\mu_\alpha cos\delta$, $\mu_\delta$(mas yr$^{-1}$)$^{**}$	 & \multicolumn{2}{c}{-0.48$\pm$0.59, -2.59$\pm$0.62} 										\\
$U, V, W$ (km s$^{-1}$)  																 & \multicolumn{2}{c}{36$\pm$1, -16$\pm$1, -18$\pm$2}									\\ 
\hline  
  \end{tabular}
\medskip

{\rm *{\em 2MASS} All-Sky Point Source Catalogue \citep{Cutri03}} 

{\rm **Newly Reduced Hipparcos Catalogue \citep{vanlee07}} \\ 
\end{table}

\section{Discussion and conclusions }
\subsection{Absolute dimensions and distance to the system}
Combination of the parameters obtained from light curves and RVs yield the absolute dimensions of the system, which are 
presented in Table\,5. The standard deviations of the parameters have been determined by JKTABSDIM\footnote{This can 
be obtained from http://www.astro.keele.ac.uk/$\sim$jkt/codes.html} code, which calculates distance and other physical 
parameters using several different sources for bolometric corrections \citep{sout05}. The mass and radii of the components 
were estimated with uncertainities of 1 \%. 

An inspection of the temperatures, masses and radii of the component stars reveals a binary system composed of two main-sequence 
stars. The temperature $T_{eff1}=19\,000$ K, mass $M_{1}=7.58$ M$_{\odot}$ and radius $R_{1}=5.8$ R$_{\odot}$ of the primary are 
consistent with the spectral type of B3, and the temperature $T_{eff2}=17\,800$ K, mass $M_{2}=5.46$ M$_{\odot}$ and radius 
$R_{2}=3.1$ R$_{\odot}$ of the secondary are consistent with an B4V spectral type star.

\begin{figure*}
  \begin{center}
      \includegraphics[scale=0.65]{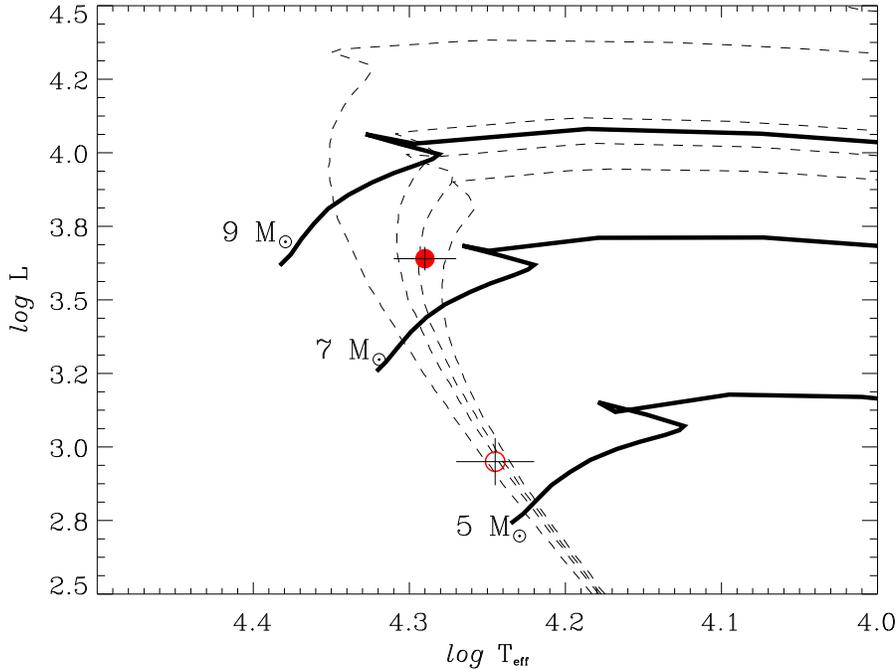}
  \end{center}
  \caption{Positions of the components of the system in HR diagram are plotted.  The solid lines evolutionary tracks for stars of various masses 
  from \citet{Scha92} and isochrones (vertical dashed lines) from \citet{Lej01} for solar metallicity with ages of 20, 25, 28, and 32 Myr going 
  from left to right. The positions of the components are consistent with an age of $\sim$ 28 Myr.
}  \label{fig:evo}
\end{figure*}

Using the two $E(B-V)$ values derived from photometric and spectroscopic data we calculated the de--reddening distance modulus
of the system. To estimate the bolometric magnitudes of the components we adopted M$_{bol}$=4.74 mag for the Sun. Using the 
bolometric corrections given by \citet{Dri00} and \citet{Gri2002} we estimate the distance to the system as 274
and 269 pc, respectively, with an uncertainity of 5 pc. However, the average distance to the system is estimated to be 
290$^{+42}_{-29}$ pc from the trigonometric parallax measured by the Hipparcos mission.

To compare the distance of HD\,194495 using a different method we used a luminosity-colour relation \citep{Bilir08}, which has 
been constructed for binary systems with main-sequence components. This method calculates the color excess ($E_{d}(B-V)$) in the 
direction and the distance of the HD\,194495 using \citet{Schlegel98} maps (see details in \citet{Bilir08}). The reduced color 
excess in direction of the HD\,194495 is calculated E$_d$(B-V)=0.072. The near-infrared magnitudes of the system 
were taken from the 2MASS Point Sources Catalogue of \citet{Cutri03} and are shown in Table 5. 

The colour excess $E(B-V)=0.08$  was estimated in direction of HD\,194495 by using equivalenth width of the interstellar lines. The near-infrared 
absolute magnitude of HD\,194495 system was estimated by the luminosity-colour relation, $M_{J}=5.228(J-H)_{o}+6.185(H-K_{s})_{0}+0.608$, of 
\citet{Bilir08} and the distance of the system is calculated as 274$\pm$8 pc by using the photometric parallaxes method. The photometric 
distance of 274$\pm$8 pc given in Table 5 is consistent with the 269$\pm$8 pc and 290$\pm$8 pc distance estimated by studiying of the 
interstellar lines and the Hipparcos measurements, respectively.

\subsection{Evolutionary stage and age of the system}
We have presented the results of the masses and radii of the stars in the HD\,194495 system to great accuracy and also a photometric
solution of the system. The results of the light and radial velocity curves analysis of the allows to derive the absolute parameters of the system. We 
have determined the masses and radii of the two stars to 1\% for the primary and secondary star.  The resulting parameters of HD\,194495 are 
given in Table\,5.  As is seen in  Table\,5, both stars of the system are well within their Roche radii but experience tidal distortion that is evident in the V-band light curve 
(Figure\,4). The rotational velocities derived from the reconstructions (Table\,3) very well match with the synchronous rotation values found 
from analysis. Therefore, this system has achieved synchronous rotation, and is not very young.

In order to discuss the evolutionary status of the components of the system, the locations of two stars were plotted on 
Hertzsprung -- Russell (H–R) diagram. In this HR diagram (Figure\,5) plotted against evolutionary tracks for stars of 
5, 7, and 9 $M_{\odot}$  from \citet{Scha92}, as well as isochrones from \citet{Lej01} for solar metallicity with ages of 
20, 25, 28, and 32\,Myr. The location of the stars is most consistent with an age of $\sim$\,28 Myr. The positions of the components 
of the system appear  in Figure\,5 to be overluminous for the derived masses of $M_{1}=7.6$ M$_{\odot}$ for the primary and 
$M_{1}=5.5$ M$_{\odot}$ for the secondary. Table\,15.7 of \citet{tokunaga00} gives the astrophysical parameters for stars 
of the various spectral classes. The mass and effective temperature of the primary fit between the listed values 
for spectral types B3 (7.6\,M$_{\odot}$) and B4 ($\sim$ 6.8\,M$_{\odot}$), but the radius is much larger than the 
means for comparable spectral types and matches a B3 star (4.8\,R$_{\odot}$ ).  
\citet{Hilditch05}  found several systems of comparable mass that, like HD\,194495, are overluminous compared with model 
predictions for eclipsing binaries in the Small Magellanic Cloud (SMC) . However, the results for HD\,194495 seem to conflict
 with the results of \citet{Malkov03}, who showed that early B-type stars that are in 
close systems and rotate more slowly than single stars are on average smaller than those same single stars.  \citet{Malkov07} also 
studied well separated binaries in an effort to use the properties of their component stars for a more 
direct comparison with single stars. The mass--luminosity--radius relations in his study, when applied to our results for HD\,194495, 
predict less-luminous, hotter, and smaller components. This is perhaps not surprising, due to the age of HD\,194495 and the 
evolution of its components from the zero-age main sequence.

To study the kinematical properties of HD\,194495, we used the system's center-of-mass' velocity, distance and proper motion values, which are given in 
Table 5. The proper motion data were taken from newly reduced Hipparcos catalogue \citep{vanlee07}, whereas the center-of-mass velocity and 
distance are obtained in this study. The system's space velocity was calculated using \citet{jonsod87} algorithm. The U, V 
and W space velocity components and their errors were obtained and given in Table 5. To obtain the space velocity precisely the first-order galactic 
differential rotation correction was taken into account \citep{mihbin81}, and -1.08 and 0.65 kms$^{-1}$ differential corrections were applied 
to U and V space velocity components, respectively. The W velocity is not affected in this first-order approximation. As for the LSR correction, 
\citet{mihbin81} values (9, 12, 7)$_{\odot}$ kms$^{-1}$ were used and the final space velocity of HD\,194495 was obtained 
as $S=43$ kms$^{-1}$. This value is in agreement with other young stars space velocities given in the  criterion from \citet{leg92} for young
disc stars  -50 $\le$ U $\le$ 20, -30 $\le$ V $\le$ 0,  -25 $\le$ W $\le$ 10.

To determine the population type of HD\,194495 the galactic orbit of the system was examined. Using \citet{Din99} N-body code, the 
system's apogalactic ($R_{max}$) and perigalactic ($R_{min}$) distances were obtained as 7.88 and 8.81 kpc, respectively. Also, the maximum 
possible vertical separation from the galactic plane is $|z_{max}|$=50 pc for the system. When determining the ellipticity the following formula 
was used:

\begin{equation}
e=\frac{R_{max}-R_{min}}{R_{max}+R_{min}}.
\end{equation}
The ellipticity was calculated as $e=0.06$. This value shows that HD\,194495 is orbiting the Galaxy in an almost circular orbit and that the system 
belongs to the young thin-disc population.

\section*{Acknowledgment}
We thank Prof.\ G.\ Strazzulla, director of the Catania Astrophysical Observatory, and Dr. G.\ Leto, responsible 
for the M. G. Fracastoro observing station for their warm hospitality and allowance of telescope time for the 
observations. This research has been also partially supported by INAF and Italian 
MIUR. This research has been made use of the ADS and CDS databases, operated at the CDS, Strasbourg, France and 
T\"{U}B\.{I}TAK ULAKB{\.I}M S\"{u}reli Yay{\i}nlar Katalo\v{g}u.


\begin{thebibliography}{00}
\bibitem[B\"{o}hm-Vitense (1981)]{Bohm81} B\"{o}hm-Vitense, E. 1981, ARA\&A, 19, 295
\bibitem[Bilir et al.(2008)]{Bilir08} Bilir S., Ak T., Soydugan E., Soydugan F., Yaz E., Ak F., Eker  Z., Demircan O. \& Helvac{\i} M. 2008, AN, 329, 835 
\bibitem[de Jager \& Nieuwenhuijzen(1987)]{JagerNie87} de Jager C. \& Nieuwenhuijzen H. 1987, A\&A, 177, 217
\bibitem[Cutri et al.(2003)]{Cutri03} Cutri R. M., et al., 2003, The IRSA 2MASS All-Sky Point Source Catalog, NASA/IPAC Infrared Science 
Archive.~http://irsa.ipac.caltech.edu/applications/Gator/ 
\bibitem[Dinescu et al.(1999)]{Din99} Dinescu, D.I., Girardi, T.M. \& van Altena, W.F. 1999, AJ, 117, 1792
\bibitem[Drilling \& Landolt(2000)]{Dri00} Drilling J. S. \& Landolt A. U., 2000, "Allen's Astrophysical Quantities", Fouth Edition, ed. A.N.Cox (Springer), p.381
\bibitem[Eggleton (1971)]{Egg71} Eggleton, P. P. 1971, MNRAS, 151, 351
\bibitem[Ekstr\"{o}m et al.(2008)]{ek08} Ekstr\"{o}m, S., Meynet, G., Maeder, A. \& Barblan, F. 2008, A\&A, 478, 467
\bibitem[Eldridge \& Tout (2004)]{Eld04} Eldridge, J. J. \& Tout, C. A. 2004, MNRAS, 353, 87
\bibitem[Girardi et al.(2002)]{Gri2002} Girardi L., Bertelli G., Bressan A., Chiosi C., Groenewegen M. A. T., Marigo P., Salasnich B.\& Weiss A. 2002, A\&A, 391, 195
\bibitem[Harmanec (1988)]{Harm88} Harmanec, P. 1988, Bull. Astron. Inst. Czech., 39, 329
\bibitem[Hern\'andez et al.(2004)]{Hernandez04} Hern\'andez J., Calvet N., Brice\~no C., Hartmann L., Berlind P. 2004, AJ, 127, 1682
\bibitem[Hilditch et al.(2005)]{Hilditch05} Hilditch, R. W., Howarth, I. D. \& Harries, T. J. 2005, MNRAS, 357, 304
\bibitem[Johnson \& Soderblom (1987)]{jonsod87} Johnson D. R. H. \& Soderblom D. R. 1987, AJ, 93, 864 
\bibitem[Kazarovets et al.(1999)]{Kazo99} Kazarovets E.V., Samus, N.N., Durlevich O.V., Kireeva N.N., \& Pastukhova E.N. 2006, IBVS, No.5721
\bibitem[Leggett (1992)]{leg92} Leggett, S. K. 1992, APJS, 82, 351 
\bibitem[Lejeune \& Schaerer(2001)]{Lej01} Lejeune, T. \& Schaerer, D. 2001, A\&A, 366, 538
\bibitem[Malkov (2003)]{Malkov03} Malkov, O. Y. 2003, A\&A, 402, 1055
\bibitem[Malkov (2007)]{Malkov07} Malkov, O. Y. 2007, MNRAS, 382, 1073
\bibitem[Mihalas \& Binney(1981)]{mihbin81} Mihalas D. \& Binney J. 1981. in Galactic Astronomy, 2nd edition, Freeman, San Fransisco, p.181
\bibitem[Monet (1979)]{Monet79} Monet,D.G. 1979, PASP, 91, 95
\bibitem[Munari \& Zwitter(1997)]{MunariZwit97} Munari U. \& Zwitter T. 1997, A\&A, 318, 269
\bibitem[Nordstr\"om et al.(1963)]{Nord63} Nordstr\"om B., Mayor M., Holmberg J., Pont F., Jorgensen B.R., Olsen E.H., Udry, S. \& Mowlavi N. 2004, A\&A, 418, 989
\bibitem[Penny et al.(2001)]{Pen01} Penny R. L., Seyle D., Gies D. R., Harvin J. A., Bagnuolo W. G. Jr, Thaller M. L., Fullerton A. W. \& Kaper L. 2001, ApJ, 548, 889
\bibitem[Pr\v{s}a \& Zwitter(2005)]{PrsaZt05} Pr\v{s}a A., Zwitter T. 2005, ApJ, 628, 426P
\bibitem[Pojmanski(2002)]{Poj02} Pojmanski G. 2002, AcA, 52, 397
\bibitem[Pols et al.(1998)]{pols98} Pols, O. R., Schroder, K.-P., Hurley, J. R., Tout, C. A., \& Eggleton, P. P. 1998, MNRAS, 298, 525
\bibitem[Queloz et al.(1998)]{Que98} Queloz D., Allain, S., Mermilliod, J.-C., Bouvier, J., \& Mayor, M., 1998, A\&A, 335, 183
\bibitem[Royer et al.(2002)]{Royer02} Royer, F., Gerbaldi, M., Faraggiana, R., \& Gomez, A. E. 2002, A\&A, 381, 105
\bibitem[Schaller et al.(1992)]{Scha92} Schaller, G., Schaerer, D., Meynet, G. \& Maeder, A. 1992, A\&AS, 96, 269
\bibitem[Schlegel et al.(1998)]{Schlegel98} Schlegel D. J., Finkbeiner D. P. \& Davis M. 1998, ApJ, 500, 525 
\bibitem[Southworth et al.(2005)]{sout05} Southworth J., Smalley B., Maxted P. F. L., Claret A. \& Etzel P. B. 2005, MNRAS, 363, 529
\bibitem[Tokunaga (1991)]{tokunaga00} Tokunaga A. T. 2000, in Cox A. N., ed., Allen's Astrophysical Quantities, 4th edn. Springer-Verlag, Berlin , p. 143
\bibitem[Williams (2009)]{Will09} Williams, S.J. 2009, AJ, 137, 3222
\bibitem[van Leeuwen(2007)]{vanlee07} van Leeuwen F. 2007, A\&A, 474, 653
\bibitem[van Hamme(1993)]{vanham03} van Hamme, W. 1993 AJ, 106, 2096 
\bibitem[von Zeipel(1924)]{Zeipel24} von Zeipel, H. 1924, MNRAS, 84, 665
\end{thebibliography}
\end{document}